\documentclass[useAMS,usenatbib,aas_macros]{mn2e}

\usepackage{epsfig}
\usepackage{amssymb}
\usepackage{natbib}
\usepackage{aas_macros}
\usepackage{amsmath}

\DeclareGraphicsExtensions{.eps}

\newcommand{\gamm}{$\gamma'$}

\newcommand{\figu}{Fig.~}
\newcommand{\figus}{Figs.~}
\newcommand{\eq}{equation~}
\newcommand{\sect}{Section~}

\newcommand{\devac}{de Vaucouleurs}

\newcommand{\re}{$R_e$}

\newcommand{\ree}{R_e}

\newcommand{\mstar}{$M_{\rm star}$}

\newcommand{\mhalo}{$M_{\rm halo}$}
\newcommand{\mhaloe}{M_{\rm halo}}

\newcommand{\msune}{M_{\odot}}

\newcommand{\mstare}{M_{\rm star}}

\newcommand{\kpce}{{\rm kpc}}

\def\papI{paper I}

\begin{document}

\voffset=-0.50in

\def\sarc{$^{\prime\prime}\!\!.$}
\def\arcsec{$^{\prime\prime}$}
\def\arcmin{$^{\prime}$}
\def\degr{$^{\circ}$}
\def\seco{$^{\rm s}\!\!.$}
\def\ls{\lower 2pt \hbox{$\;\scriptscriptstyle \buildrel<\over\sim\;$}}
\def\gs{\lower 2pt \hbox{$\;\scriptscriptstyle \buildrel>\over\sim\;$}}

\title[Dependence of \gamm\ on redshift]{Revisiting the Bulge-Halo Conspiracy II: Towards explaining its puzzling dependence on redshift}

\author[F. Shankar et al.]
{Francesco Shankar$^{1}$\thanks{E-mail:$\;$F.Shankar@soton.ac.uk},
Alessandro Sonnenfeld$^{2,3,4}$,
Philip Grylls$^{1}$, Lorenzo Zanisi$^{1,5}$, \newauthor Carlo Nipoti$^{6}$, Kyu-Hyun Chae$^{7}$, Mariangela Bernardi$^{8}$,
Carlo Enrico Petrillo$^{9}$, \newauthor Marc Huertas-Company$^{10,11}$, Gary A. Mamon$^{12}$, Stewart Buchan$^{1}$
\\
$1$ Department of Physics and Astronomy, University of Southampton, Highfield, SO17 1BJ, UK\\
$2$ Kavli IPMU (WPI), UTIAS, The University of Tokyo, Kashiwa, Chiba 277-8583, Japan\\
$3$ Physics Department, University of California, Santa Barbara, CA 93106, USA\\
$4$ Physics and Astronomy Department University of California Los Angeles CA 90095-1547\\
$5$ Dipartimento di Fisica, Universit\'{a} di Roma ``La Sapienza'', P.le Aldo Moro 2, 00185, Roma, Italy\\
$6$ Department of Physics and Astronomy, Bologna University, via Gobetti 93/2, I-40129 Bologna, Italy\\
$7$ Department of Physics and Astronomy, Sejong University, 209 Neungdong-ro Gwangjin-Gu, Seoul 05006, Korea\\
$8$ Department of Physics and Astronomy, University of Pennsylvania, 209 South 33rd St, Philadelphia, PA 19104\\
$9$ Kapteyn Astronomical Institute, University of Groningen, Postbus 800, 9700 AV, Groningen, The Netherlands\\
$10$ LERMA, Observatoire de Paris, CNRS, Universit\'{e} Paris Diderot, Paris Sciences et Lettres (PSL) Research University, Universit\'{e}\\
$11$ Universit\'{e} Paris Denis Diderot, 75205 Paris Cedex 13, France\\
$12$ Institut d'Astrophysique de Paris (UMR 7095: CNRS \& UPMC, Sorbonne Universit\'{e}s), 98bis Bd Arago, F-75014 Paris, France}
\date{}
\pagerange{\pageref{firstpage}--
\pageref{lastpage}} \pubyear{2017}
\maketitle
\label{firstpage}

\begin{abstract}
We carry out a systematic investigation of the total mass density profile of massive ($\log \mstare/\msune \sim 11.5$) early-type galaxies and its dependence on redshift,
specifically in the range $0\lesssim z \lesssim 1$. We start from a large sample of SDSS early-type galaxies with stellar masses and effective radii measured assuming two different profiles,
\devac\ and S\'{e}rsic. We assign dark matter haloes to galaxies via abundance matching relations with standard $\Lambda$CDM profiles and concentrations.
We then compute the total, mass-weighted density slope at the effective radius \gamm, and study its redshift dependence at fixed stellar mass.
We find that a necessary condition to induce an increasingly flatter \gamm\ at higher redshifts, as suggested by current strong lensing data, is to allow the intrinsic stellar profile of massive galaxies to be S\'{e}rsic and the input S\'{e}rsic index $n$ to vary with redshift as $n(z)\propto (1+z)^{\delta}$, with $\delta \lesssim -1$. This conclusion holds irrespective of the input \mstar-\mhalo\ relation, the assumed stellar initial mass function, or even the chosen level of adiabatic contraction in the model. \emph{Secondary} contributors to the observed redshift evolution of \gamm\ may come from an increased contribution at higher redshifts of adiabatic contraction and/or bottom-light stellar initial mass functions. The strong lensing selection effects we have simulated seem not to contribute to this effect. A steadily increasing S\'{e}rsic index with cosmic time is supported by independent observations, though it is not yet clear whether cosmological hierarchical models (e.g., mergers) are capable of reproducing such a fast and sharp evolution.
\end{abstract}


\section{Introduction}
\label{sec|intro}

The details of the evolution of the most massive galaxies are still a matter of debate. One of the most promising scenarios is the so-called two-phase evolutionary sequence, in which a fraction of the stars is created ``in-situ'', and the rest forms ``ex situ'' and is accreted at later times. It is however still quite debated what the relative contribution actually is between the fraction of stars formed in the initial burst of star formation, and the fraction of stars acquired on cosmological timescales via, e.g., mergers and/or cosmic flows \citep[e.g.,][]{Ciras05,Lapi06,Shankar06,Delucia07,Hop08clust,Dekel09b,Naab09,Oser10,ShankarRe,ShankarPhire,Gonzalez11,Lapi11,Hirschmann13,Shankar13,Aversa15,BuchanShankar}.

Valuable additional observational constraints on the evolutionary patterns of massive galaxies are derived from the structural evolution across cosmic time and galaxy properties. In particular, thanks to a variety of data from strong lensing and stellar dynamics, the last decade has seen a dramatic boost in the interest towards the ``total'' mass density profile of massive galaxies.
The latter, on average well approximated by a power-law $\rho(r)\propto r^{-\gamma}$ with slope $\gamma \approx 2$
\citep[see, e.g.,][]{TreuKoopmans04,Koop06,Gavazzi07,Koop09,Barnabe11}, has revealed clear trends with stellar mass, effective radius, and redshift \citep[e.g.,][]{Auger10,Ruff11,Bolton12,Sonne13,Dye14,Tortora14a,Tortora14b}.
The slope decreases for increasing galaxy size at fixed stellar mass, and for lower mass galaxies at fixed effective radius.

Building on previous work \citep[e.g.,][]{Newman13b,DuttonTreu,Newman15}, \citet[][\papI, hereafter]{Shankar17g} devised accurate semi-empirical models to create advanced mock galaxy catalogues based on a large galaxy sample with measured effective radii and stellar masses, together with basic assumptions on their host dark matter haloes (abundance matching). By varying the stellar mass profile, stellar initial mass function (IMF), stellar mass-halo mass relation, they concluded that the observed variations of the mass density slope with galaxy properties is naturally explained by the interplay between the combined stellar and halo profiles, with a preference for heavy stellar IMFs, and for uncontracted dark matter halo profiles as those expected from dark matter-only cosmological N-body simulations, at least for galaxies not in cluster environments.

Regarding cosmic time evolution, in particular, the total mass density slope has been observed to anti-correlate with redshift, being shallower for higher-redshift objects with respect to local systems of the same mass and size \citep[e.g.,][]{Ruff11,Bolton12,Sonne13,Dye14}. Explaining this trend is the focus of this second paper in the series. As the total mass density distribution is the sum of the density distributions of the stars and of the dark-matter halo,
its gradual variation along cosmic time is expected to encode important clues on the relative evolution of these two components.
In turn, being able to identify the key assumptions in the semi-empirical models that can best reproduce the observational
data will shed light on the relative roles played by different processes in shaping galaxies.
For example, if (dry) mergers are truly gradually growing the sizes of galaxies at late epochs, then one would naively expect somewhat \emph{steeper} values of $\gamma$
for higher-redshift, more compact, massive galaxies, a trend which could be directly tested in our semi-empirical models.

The structure of this work is the following.
In Section \ref{sec|EmpiricalModel} we briefly recap the semi-empirical model used to construct mock samples of galaxies.
In Section \ref{sec|Results} we present a series of models of the total mass density slope as a function of redshift, mainly focused on variations of the S\'{e}rsic index and stellar IMF and dark matter fractions, which we compare with strong lensing observations.
In Section \ref{sec|discu} we discuss our results in view of evolution of the mass density slope along the putative progenitors and other issues of interest, and conclude in Section \ref{sec|Conclu}.

In the following we will adopt a reference cosmology
with parameters
$\Omega_{\rm m}=0.30$, $\Omega_{\rm b}=0.045$, $h=0.70$,
$\Omega_\Lambda=0.70$, $n=1$, and $\sigma_8=0.8$, to match those usually
adopted in the observational studies on the stellar
mass function and strong lensing considered in this work.

\section{The Data}
\label{sec|Data}

Observations of the total mean mass density slope \gamm\ are taken from the gravitational strong lensing analysis of \citet{Sonne13}.
The latter combined strong lensing and stellar kinematics data for 84 lens galaxies in the redshift interval $0.1 < z < 1$.
They constrained the density slope of each object by fitting a spherical power-law density profile to the observed Einstein radius and central velocity dispersion.
The model velocity dispersion was computed by solving the spherical Jeans equation under the assumption of isotropic orbits.
\citet{Sonne13} then fitted for the distribution of the slope \gamm\ across the population of lenses.
They described the distribution of \gamm\ as a Gaussian with a mean that is a function redshift, stellar mass and size, given by

\begin{equation}
\langle \gamma' \rangle=\gamma_0+\alpha(z-0.3)+\beta(\log \mstare-11.5)+\xi \log (R_{\rm e}/5) \, ,
\label{eq|Sonne13gamma}
\end{equation}
with $\gamma_0=2.08^{+0.02}_{-0.02}$, $\beta=0.40^{+0.16}_{-0.15}$, and $\xi=-0.76^{+0.15}_{-0.15}$, with an average dispersion around the median of
$\sigma_{\gamma'}=0.12^{+0.02}_{-0.02}$.

Effective radii of the lens galaxies were obtained by fitting \devac\ profiles. Stellar masses were measured through stellar population synthesis, using \devac\ magnitudes as input, and assuming a Salpeter IMF. Of particular relevance to this work, the constrained value for the redshift evolution parameter $\alpha=-0.31^{+0.09}_{-0.10}$, at face value would imply that, at fixed stellar mass and effective radius, a null evolution of \gamm\ with redshift is rejected at $\gtrsim 3\sigma$ level. This trend with redshift was also anticipated by \citet{Bolton12} and \citet{Ruff11}, and later also confirmed by \citet{Dye14} who, analyzing five lens systems in the range $0.22<z<0.94$, concluded that the logarithmic slope of the total mass density profile steepens with decreasing redshift. \citet{Dye14} also support a decreasing \gamm\ with increasing effective radius at fixed redshift, in full agreement with \eq\ref{eq|Sonne13gamma} (and the modelling put forward in \papI).

To create the mock galaxy catalogue to compare to the \citet{Sonne13} strong lensing data, we start from the \citet{Meert15} sample extracted from the Sloan Digital Sky Survey (SDSS) DR7 spectroscopic sample \citep{2009ApJS..182..543A} in the redshift range $0.05<z<0.2$, and with a probability $P(E)>0.85$ of being elliptical galaxies based on the Bayesian automated morphological classifier by \citet{Huertas11}. For this SDSS subsample we have at our disposal
three different light profiles, a pure \devac\ \citep{deVac}, a S\'{e}rsic+Exponential, and a pure S\'{e}rsic \citep{Sersic63} profile
\citep[see, e.g.,][]{Bernardi13,Bernardi17}. Slightly more accurate mass-to-light ratios from \citep{Mendel13} have been adopted for this sample with respect to those considered in \papI, which were based on \citet{Bell03SEDs}. We verified that all the relevant scaling relations in terms of (\devac) stellar mass, size, and velocity dispersion are in equally good, if not even improved, agreement to the \citet{Sonne13} sample.

To probe for a possible redshift evolution of the total galaxy mass density profiles, we will create redshift-dependent mock galaxy catalogues starting from our $z<0.2$ SDSS galaxy sample and include
some structural evolution in their light profiles as inspired by observational results. This is a safe approximation given that the number densities and morphological mix of galaxies have a weak dependence on redshift up to $z\lesssim 1$ \citep[see, e.g.,][their \figu3]{Huertas15}. In what follows we will thus always assume that the effective radii of massive galaxies decrease with increasing redshift as $\ree \propto (1+z)^{-1}$ at fixed stellar mass, in line with a number of independent observations of massive galaxies at high redshifts \citep[e.g.,][and references therein]{Cimatti12,Huertas13a,van14}.
To isolate the redshift evolution in \gamm\ from its dependence on stellar mass and effective radius, which we have already extensively explored in \papI, in \sect\ref{sec|Results} we will mostly focus on galaxies in very narrow bins of (\devac) stellar mass $11.4<\log \mstare<11.6$ and corresponding effective radii at $z\lesssim 0.1$ in the range $4<\ree/\kpce<6$. These represent the characteristic masses and sizes of typical galaxies in the reference observational sample of \citet{Sonne13}, as also evident from \eq\ref{eq|Sonne13gamma}.

The main difference between this work and \papI\ is that in the latter we varied the assumed \emph{intrinsic} stellar profile of our mock galaxies according to the choice of light profile adopted in the observational sample we were comparing with, i.e., \devac\ when comparing with \citet{Sonne13} and S\'{e}rsic when comparing with \citet{Newman13b,Newman15}. Here we further expand on this methodology by distinguishing between an intrinsic and observed profile, with the former being what we believe is a more realistic modelling of the stellar profile of typical massive galaxies, and the latter being the actual one assumed in the observational sample used to compare the models with. In this logic, stellar masses and effective radii in this work will always be quoted as \devac\ quantities, as we will only compare to \citet{Sonne13}, but the intrinsic ones will vary from \devac\ to S\'{e}rsic profiles.

\section{The Semi-Empirical Model}
\label{sec|EmpiricalModel}

We here briefly describe the main features of the semi-empirical model adopted in this work, leaving the full details to \papI, and mostly highlighting the relevant differences/improvements with respect to it.

The nearly-isothermal total mass density profile inferred from strong lensing and stellar dynamics \citep[e.g.,][]{Mamon05b,Sonne13,Cappellari15} is believed to be a result
of the combined steeper and flatter mass density profiles of, respectively, the stellar and dark matter components (known as the ``bulge-halo conspiracy''). As extensively discussed in \papI, the relative contributions of these components within scales comparable to the half-mass radii of the galaxies should then dictate the relative trends of the mass density slope with large-scale galaxy properties such as stellar mass and effective radius.

In our galaxy mocks, halo masses are assigned to each galaxy in our SDSS sample via abundance matching techniques. More precisely, at any redshift of interest we compute the mean relation between stellar mass and host halo mass via the equality between the cumulative stellar and (sub)halo mass functions
\begin{equation}
\Phi(>\mstare,z)=\Phi_h(>\mhaloe,z) \, .
\label{eq|Cum}
\end{equation}
We include, throughout, a scatter of $0.15$ dex in stellar mass at fixed halo mass constant with halo mass and redshift, a good approximation at these mass scales and at redshifts $z\lesssim 1$ \citep[e.g.,][]{Shankar14b}. We incorporate scatter in the abundance matching by refitting the high mass-end slope of the scatter-free \mstar-\mhalo\ relation (\eq9 in \papI) until it matches the input stellar mass function. As detailed in \papI\ we then invert the \mstar-\mhalo\ relation by self-consistently computing the implied mean and scatter in halo mass as a function of stellar mass.

For the dark matter component in the right-hand side of \eq\ref{eq|Cum} we adopt the host halo mass function of \citet{Tinker08}, with (unstripped) subhalo correction from \citet{Behroozi13}. Halo masses $\mhaloe$ are always defined within $r_{200c}$, such that the average density within $r_{200c}$ is 200 times the critical density of the Universe at redshift $z$. We always assume a \citet{NFW} profile which, as proven in \papI, is a good approximation for (central) galaxies at the mass scale $\log \mstare/\msune \sim 11.5$ of interest to this work. The concentration-halo mass relation is taken from \citet{Bene14} with a log-normal scatter of $0.16$ dex. We also consider NFW profiles modified by the inclusion of adiabatic contraction or adiabatic expansion following the methodology outlined by, e.g., \citet[][]{Dutton07} and \citet[][]{Barausse12}. Briefly,
the final and initial radius of the mass distribution are related by the expression $r_f=\Gamma^{\nu}r_i$, with $\Gamma$ the contraction factor \citep{Blumenthal}, and the parameter $\nu$ taking on the values $\nu=0,0.8,-0.3$ for no contraction, adiabatic contraction, and adiabatic expansion, respectively (full details in \papI).

For the stellar component on the left-hand side of \eq\ref{eq|Cum} we rely on the \citet{Bernardi17} stellar mass function. The latter has a specific shape which we self-consistently vary according to the assumed stellar profile (\devac or S\'{e}rsic), and/or assumed IMF (see below), before applying \eq\ref{eq|Cum}.
S\'{e}rsic-based profiles tend to produce substantially higher number densities of massive galaxies and thus steeper \mstar-\mhalo\ relations via \eq\ref{eq|Cum} (see, e.g., \figu1 in \papI). For the remainder of this paper we assume each rendition of the \citet{Bernardi17} stellar mass function to be strictly constant in the redshift range $0<z<1$. Forcing a constant number density of massive galaxies up to $z\sim 1$ may be an oversimplification. Recent abundance matching and clustering modelling tends, in fact, to favour some stellar mass growth in the most massive galaxies (Buchan et al. 2017, submitted). This in turn would necessarily induce some late evolution in the high mass-end of the stellar mass function, thus impacting the exact shape of the implied stellar mass-halo mass relation \citep[e.g.,][]{Kravtsov14,Shankar14b,Skibba14,Coupon15,Patel15,Rodriguez14,Shan15}. As further discussed below, however, the exact choice of input stellar mass function and/or \mstar-\mhalo\ relation does not alter any of our conclusions, in the redshift range $0<z<1$. This is mainly because the dark matter fraction within the effective radius in massive galaxies remains relatively contained to $\lesssim 40\%$ even at high redshifts \citep[see, e.g.,][and \sect\ref{subsec|DMfracRedshift}]{WM13,Shetty15}, coupled to a steep central stellar profile.

For each light profile we explore the impact of varying the input IMF, from a Chabrier \citep{Chabrier03}, the reference one adopted by \citet{Bernardi17}, to a Salpeter \citep{SalpeterIMF}, and
variable (velocity dispersion-dependent) IMF. For the former, we simply add 0.25 dex to each stellar mass \citep[][]{Bernardi10}. For the latter,
we follow \citet{Cappellari15} and \citet[][their \figu19]{Cappellari16}, and express the variation in mass-to-light ratio with respect to Salpeter as
\begin{eqnarray}
\begin{aligned}
\log \left(M/L\right)[\sigma_e]=\log \left(M/L\right)_{\rm Salp}+\rm{f}(\sigma_e)=\\
\log \left(M/L\right)_{\rm Salp}-0.0576+0.364\times \log \left(\frac{\sigma_e}{200\,\, \rm{km\, s^{-1}}}\right)
\label{eq|MLratioCappellari}
\end{aligned}
\end{eqnarray}
with an intrinsic scatter of 0.11 dex. We note that in this paper we will mainly focus on galaxies with stellar masses peaked around $\log \mstare/\msune \sim 11.5$ (Salpeter IMF), for which \eq\ref{eq|MLratioCappellari} yields stellar masses $\sim 0.20$ dex higher than a Chabrier IMF.
We will thus not be able to fully probe the validity of \eq\ref{eq|MLratioCappellari} to all stellar masses, though we will be able to discern if a {\bf heavy} IMF as the one given in \label{eq|MLratioCappellari} is more appropriate than a Chabrier IMF in matching the data. We will also consider a model characterized by a variable IMF as in \eq\ref{eq|MLratioCappellari}, which gradually tends towards a Chabrier IMF linearly with redshift as
\begin{eqnarray}
\log \left(M/L\right)[\sigma_e,z]=\log \left(M/L\right)_{\rm Chab}+\left[0.25+\rm{f}(\sigma_e)\right](1-z)
\label{eq|MLratioCappellariRedshift}
\end{eqnarray}
where we have assumed that a Chabrier IMF induces 0.25 dex lower stellar masses than a Salpeter IMF \citep[e.g.,][]{Bernardi10}. \eq\ref{eq|MLratioCappellariRedshift} clearly reduces to \eq\ref{eq|MLratioCappellari} at $z=0$. We note that the gradual transformation with redshift proposed in \eq\ref{eq|MLratioCappellariRedshift}, even though possibly not fully aligned with other proposals in the literature \citep[see, e.g.,][]{Sonne17}, is the only one still acceptable by our semi-empirical modelling, as it will become apparent in \sect\ref{subsec|EvolutionIMF}.
A series of studies in recent years have also suggested that the IMF of massive early-type galaxies should be heavier
in the centre and lighter in the outskirts \citep{Martin15b, Martin15c, LaBarbera16, Conroy17, vanDokkum17}.
Following \citet{vanDokkum17}, for completeness we also present a model with a scale-dependent IMF parameterized\footnote{We note that
the normalization of \eq\ref{eq|MLRvanDokkum} has been reduced by $\sim 10\%$, still within the range allowed by \citet{vanDokkum17}'s data.
Any higher normalization would further exacerbate the poor match between this model and the data reported in \figu\ref{fig|GammaEvolutionSer}.
Also, \eq\ref{eq|MLRvanDokkum} has been strictly derived as a function of \emph{projected} radii, though we will here simply extend it to 3D mass density distributions.} as
\begin{equation}
    \Theta\left[R/\ree\right]=
\begin{cases}
    \log[2.23-3.6(R/\ree)],& \text{if } R/\ree<0.4\\
    0\, ,              & \text{otherwise}
\end{cases}
\label{eq|MLRvanDokkum}
\end{equation}
where $\Theta\left[R/\ree\right]=\log \left(M/L\right)\left[R/\ree\right]-\log \left(M/L\right)_{\rm MW}$.
We verified that \eq\ref{eq|MLRvanDokkum} implies an average
increase in stellar mass with respect to a Milky Way-type IMF \citep{Kroupa01} of $\sim 0.04$ dex, with negligible dependence on stellar mass or effective radius. We
then further add 0.05 dex to convert from a Kroupa \citep{Kroupa01} to our reference Chabrier IMF \citep[e.g.,][]{Bernardi10}.

\begin{figure}
    \center{\includegraphics[width=8.5truecm]{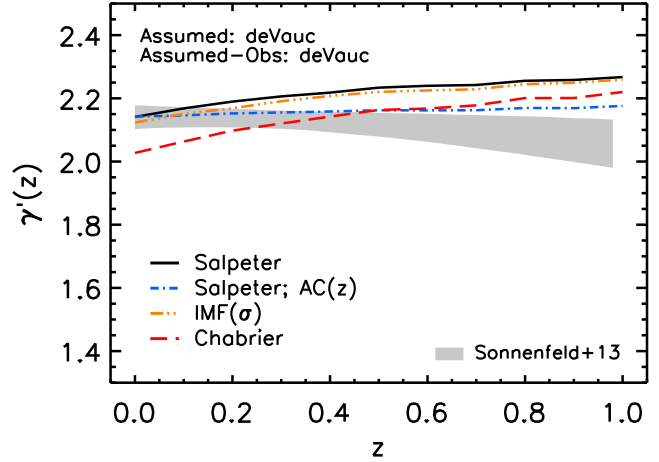}
    \caption{{Predicted redshift evolution of the slope of the total mean mass density profile, measured at the effective radius, \gamm,
    for galaxies with stellar masses in the narrow interval $11.4<\log \mstare/\msune<11.6$ and effective radius at $z\le 0.1$ in the range $4<\ree(z)/{\rm kpc}<6$, steadily decreasing proportionally to $(1+z)^{-1}$ at higher redshifts.
    All models assume an intrinsic \devac\ profile, as in the observations, and a $\propto (1+z)^{-1}$ redshift evolution in the effective radius \re, but different IMFs, as labelled. The orange, triple dot-dashed line also includes an increasing adiabatic contraction towards higher redshifts (see text for details). Models are compared with data from \citet{Sonne13} (grey band).
    \label{fig|GammaEvolutionDeVauc}}}}
\end{figure}

\begin{figure*}
    \center{\includegraphics[width=15truecm]{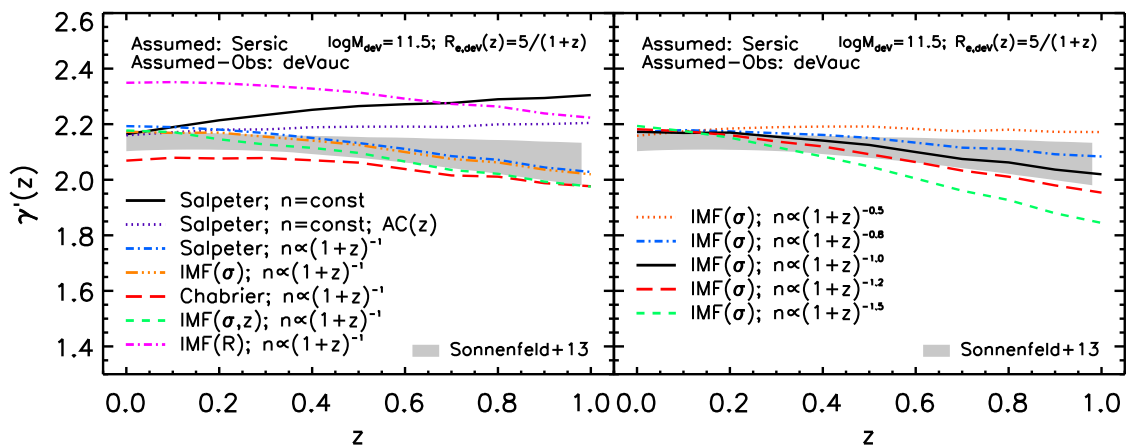}
    \caption{Same format as \figu\ref{fig|GammaEvolutionDeVauc}. Left: Comparison of models that assume an intrinsic S\'{e}rsic stellar profile
    and observed \devac\ stellar profile, a $(1+z)^{-1}$ redshift evolution in the effective radius \re, but different IMFs, as labelled.
    The purple, dot-dashed line also includes an increasing adiabatic contraction towards higher redshifts (see text for details).
    Right: Comparison of models with intrinsic S\'{e}rsic
    stellar profile and different degree of redshift evolution in the S\'{e}rsic index, as labelled.
    Only models with strong redshift evolution in the S\'{e}rsic index, $n\propto (1+z)^{-\delta}$ with $\delta\gtrsim 0.7$, can broadly line up with the data.
    \label{fig|GammaEvolutionSer}}}
\end{figure*}

Finally, for each galaxy in our mock we compute the mass-weighted slope of the total density profile within a radius $r$ \citep{DuttonTreu}
\begin{eqnarray}
\begin{aligned}
\gamma'(r)&\equiv-\frac{1}{M(<r)}\int_{0}^{r}\frac{d\,{\rm log}\, \rho}{d\,{\rm log}\, x}4\pi x^2\rho(x)dx=\\  
&=3-{\rm \left. \frac{\emph{d}\,{\rm log} \, M}{\emph{d}\, {\rm log}\, \emph{x}} \right|_{\emph{x}=\emph{r}}}\, .
\label{eq|gamma}
\end{aligned} 
\end{eqnarray}
\citet[][cfr. their \figu6]{Sonne13} and \citet{DuttonTreu} have shown that
the mass-weighted slope \gamm\ computed within the effective radius \re,
is a good proxy of \gamm\ derived from strong lensing and dynamical measurements \citep[see][for a full discussion ]{Sonne13}.
In each model \gamm\ will be self-consistently computed at the \devac\ effective radius, though we note that \gamm\ has anyway a very weak dependence on scale within the range $(0.5-1.2)\ree$. Measuring \gamm\ at, say, the S\'{e}rsic effective radius instead, which could be up to a factor of $\lesssim 2$ higher than the \devac\ effective radius (cfr. \figu1 in \papI), could induce an appreciable systematic drop of $\lesssim 10\%$ in the predicted \gamm. However, this would,
if anything, exacerbate the tensions between model predictions and data.

\section{Results}
\label{sec|Results}

\subsection{Constraints on the stellar profile: An evolving S\'{e}rsic index}
\label{subsec|EvolutionSersicIndex}

\figu\ref{fig|GammaEvolutionDeVauc} shows\footnote{To be fully consistent with \papI, specifically for \figu1 (only!), we continue adopting
the mass-to-light ratio of \citet{Bernardi13}, and a \devac+Exponential (the so-called ``\texttt{cmodel}'' magnitudes) total stellar mass function to include in the abundance matching.} the predicted evolution in \gamm\ for galaxies in very narrow bins of (\devac) stellar mass $11.4<\log \mstare<11.6$ and effective radius at $z=0.1$ of $4<\ree(z)/\kpce<6$, steadily decreasing proportionally to $(1+z)^{-1}$ at higher redshifts.
A number of variants are reported in \figu1, with models characterized by a Salpeter, a variable, and a Chabrier IMF, as labelled. 
The grey area in \figu\ref{fig|GammaEvolutionDeVauc} represents the median and 68\% percentiles of the full posterior
probability distribution of \gamm\ inferred by \citet{Sonne13} for galaxies at fixed stellar mass and redshift-dependent size.
More precisely, at each redshift of interest we evaluate \gamm\ from \eq\ref{eq|Sonne13gamma} for $\log \mstare/\msune=11.5$ and $\ree(z)=5/(1+z)\, \kpce$, propagating the uncertainties on the parameters $\alpha$ and $\gamma_0$. The resulting \gamm\ appears rather constant, or at most slightly flattening, with increasing redshift. This behaviour is expected from \eq\ref{eq|Sonne13gamma} which predicts a substantial decrease of \gamm\ at higher redshifts, partly compensated by the steep (inverse) dependence on effective radius, the latter assumed to steadily decrease at earlier cosmic epochs.

As anticipated above, we choose very narrow intervals of stellar mass and effective radius for both the models and the data to highlight the dependence on redshift. It is important to stress that none of the main results discussed in this \sect\ relies on the exact choice of bins in stellar mass and/or effective radius chosen for the analysis, as long as the same selections are self-consistently applied to both the models and the strong lensing data (via \eq\ref{eq|Sonne13gamma}).

\begin{figure*}
    \center{\includegraphics[width=15truecm]{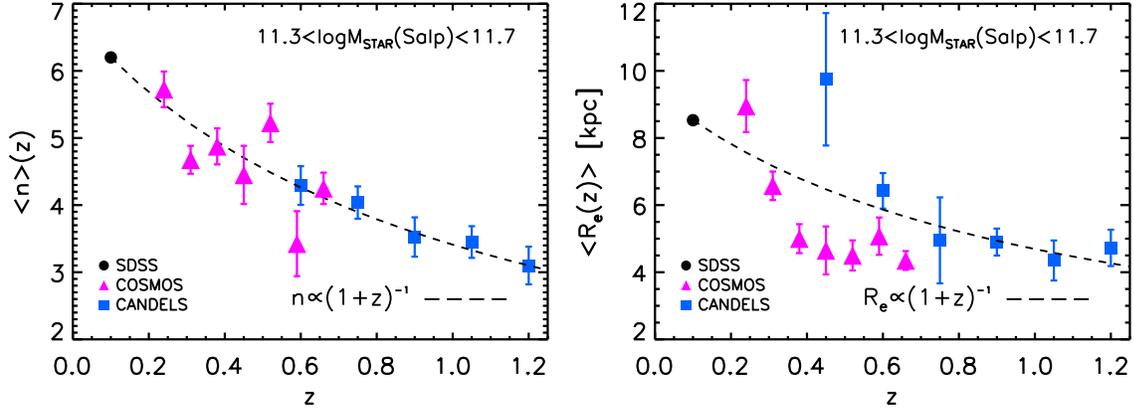}
    \caption{Measured median S\'{e}rsic index $n(z)$ (left panel) and effective radius \re$(z)$ (right panel) as a function of
    redshift from COSMOS \citep{Huertas13a} and CANDELS \citep{Huertas15}, for galaxies at constant stellar mass in the
    range $11.3<\log \mstare/\msune <11.7$. The filled, black circle at $z=0.1$ is from SDSS \citep{Bernardi12}.
    All the data are consistent with both quantities decreasing as $(1+z)^{-1}$ (dashed, black lines). \label{fig|SersicIndexRedshift}}}
\end{figure*}

In line with what inferred in \papI, we confirm that when comparing with the \citet[][grey band]{Sonne13} data at $z\lesssim 0.2$, a Chabrier IMF appears to be disfavoured with respect to a variable or Salpeter IMF. What is new and more relevant to this specific work, is that all models predict at earlier cosmic epochs a flat, if not even slightly increasing, \gamm. This (constant) behavior in the mean mass density profile is expected at some level as both the input galaxy stellar mass and effective radius are assumed fixed in time. Moreover, abundance matching predicts host halo masses to vary weakly at these galaxy mass scales, and this holds true either assuming a constant or slightly decreasing number density of massive galaxies up to $z\sim 1$, as anticipated in \sect\ref{sec|EmpiricalModel} \citep[e.g.,][]{Shankar06,Moster13,Shankar14b,Bernardi16,Tinker16}. For instance, we checked that
very similar results to those reported in \figu\ref{fig|GammaEvolutionDeVauc}
are obtained assuming a strictly non-evolving stellar mass-halo mass relation instead,
or switching to the \citet{Moster13} relation, both of which inherently assume some degree of evolution in the high-mass end of the stellar mass function.
Assuming an ad-hoc increase in the contribution of the (flatter)
dark matter distribution within the effective radius, via for example a steadily increasing AC up to $z\sim 1$ (dot-dashed, blue line), tends to induce a decrease in \gamm, better aligning model outputs with the strong lensing data, though still falling short in fully matching the data at $z\gtrsim 0.7$. Despite the apparent improvement, a substantial contribution of AC is in general disfavoured by the data, as emphasized in \papI\ and further discussed below.

At this point, despite having explored all the possible combinations of the input parameters (more relevantly IMF and host halo mass), we still fall short in reproducing the steep decline in \gamm\ at higher redshifts. It is clear that additional assumptions/parameters are required by the data. This is a perfect example of the power and usefulness of the semi-empirical approach. The least possible assumptions and associated parameters are initially included in the models. Gradually additional degrees of complexities can be included in the model, wherever needed. This allows for extreme flexibility and transparency, avoiding the risk of being clouded by an initially too heavy modelling.

In what follows, unless otherwise noted, for the stellar component we will focus on S\'{e}rsic profiles. The latter have proven to be a much more suitable model\footnote{More specifically, \citet{Bernardi12} have shown that a S\'{e}rsic+Exponential profile is the most appropriate light profile model for SDSS galaxies. However, the latter is not ideally suited to compute total mass density profiles as given by \eq\ref{eq|gamma} which strictly assumes spherical symmetry. The disc component is anyway relatively small in the stellar mass scale of galaxies of interest here.} to properly capture the full extent and variety of the light distribution in especially massive galaxies
\citep[e.g.,][and references therein]{Bernardi12}. Thus we will first compute the total mass density slope \gamm\ competing to each galaxy in our sample assuming an \emph{intrinsic} S\'{e}rsic profile for the stellar component (and NFW for the dark matter one). However, all our predictions on \gamm\ will continue being expressed in terms of their \devac\ counterparts in stellar mass and effective radius (available from our SDSS sample, see \sect\ref{sec|Data}), to properly compare to the strong lensing observations by \citet{Sonne13} who have always assumed a \devac\ profile. We still continue assuming that all half-mass radii (for \emph{both} the assumed intrinsic and observed stellar profiles) evolve as $\propto (1+z)^{-1}$.

\figu\ref{fig|GammaEvolutionSer} compares the predictions of a variety of models with intrinsic S\'{e}rsic stellar profiles and \devac\ stellar masses and effective radii within the intervals $11.4<\log \mstare<11.6$ and $4<\ree/\kpce<6$ at $z\leq0.1$, respectively, as in the observations (grey bands).
These types of light profiles effectively add a degree of freedom to our semi-empirical models, namely the S\'{e}rsic index $n$. As shown in Appendix B of \papI,
increasing $n$ at fixed stellar mass and size produces an overall steeper stellar profile and thus naturally induces a larger total mass density slope \gamm.
Irrespective of the amount of adiabatic contraction, the models characterized by a constant S\'{e}rsic index $n$ with time (black solid and purple dotted lines in the left panel), inevitably predict a more or less steep dependence of \gamm\ on redshift, at variance with the data but in line with \figu1.
This would hold irrespective of the choice of (time-independent) IMF, which would simply shift the black line vertically but without erasing its strong
dependence on redshift (see also \figu\ref{fig|GammaEvolutionDeVauc}).

Assuming instead a decrease of the S\'{e}rsic index with increasing redshift, as $n\propto (1+z)^{-1}$, tends to flatten the stellar profile at fixed stellar mass and effective radius, thus inducing an overall lower \gamm. The latter behavior holds irrespective of the input IMF, Salpeter, variable in velocity dispersion, variable in velocity dispersion and redshift, scale-dependent, Chabrier
(blue dot-dashed, orange triple dot-dashed, green dashed, magenta dot-dashed, and red long-dashed lines, in the left panel of \figu\ref{fig|GammaEvolutionSer}). A scale-dependent IMF, in particular, yields a significantly steeper \gamm\ with respect
to the data, even in the presence of maximal AC, which, we verified would simply reduce \gamm\ by $\sim 25\%$.
This behaviour is expected given
that the stellar profile gets more prominent in the inner regions (\eq\ref{eq|MLRvanDokkum}).

The right panel of \figu\ref{fig|GammaEvolutionSer} shows instead the same model with
velocity dispersion-dependent IMF, and different degrees of evolution in the S\'{e}rsic index with redshift, $n\propto (1+z)^{-\delta}$.
A $\delta\sim 0.5$ model is already sufficient to
fully erase the dependence of \gamm\ on redshift
(dotted, orange line), while any stronger dependence characterized by $\delta \gtrsim 0.7$ can reverse the trend of \gamm\ with redshift. On the other hand,
too strong variations of the S\'{e}rsic index with time beyond $\delta \gtrsim 1.2$, tend to be disfavoured by the data, at least when the selection is made
at fixed stellar mass (but see also \sect\ref{subsec|Progenitors}).
All in all, we conclude that a substantial variation in the S\'{e}rsic index with redshift of the type $n\propto (1+z)^{-\delta}$, with $0.5 \lesssim \delta \lesssim 1.2$,
is a necessary condition to align models with the strong lensing data examined in this work.

\begin{figure*}
    \center{\includegraphics[width=18truecm]{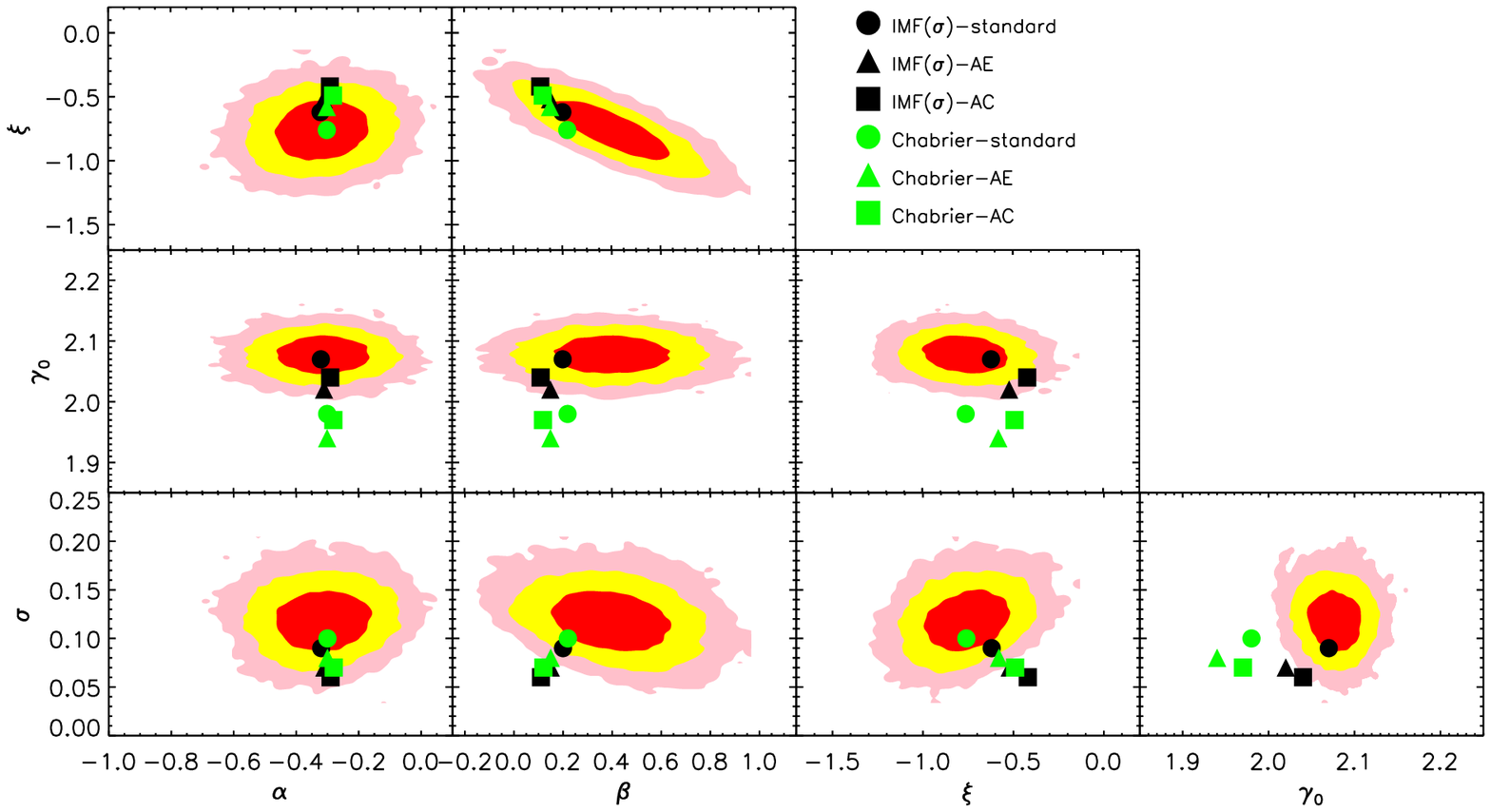}
    \caption{Covariances of posterior probability distributions of pairs of model parameters of \eq\ref{eq|Sonne13gamma}.
    The red, yellow, and pink regions mark the 1$\sigma$, 2$\sigma$, and 3$\sigma$ contour levels for each pair of
    the parameter space extracted from the \citet{Sonne13} data. The data are compared with the predictions from the
    semi-empirical models based on a variable (\eq\ref{eq|MLratioCappellari}) and Chabrier Initial Mass Functions
    (black and green symbols, respectively). The circles, triangles, and squares refer, respectively, to models
    with standard profiles, with adiabatic expansion, and with adiabatic contraction, as labelled. \label{fig|GammaPosteriorProbIMF}}}
\end{figure*}

On the observational side, the issue of the structural evolution of massive galaxies in terms of size, S\'{e}rsic index and (central) surface mass density has been a subject of hot debate in the last years \citep[e.g.,][]{Ascaso13}. At both fixed stellar mass or cumulative number density, effective radii and S\'{e}rsic indices have been claimed to evolve with redshift \citep[e.g.,][]{vandokkum10,Mori14,Huertas15}. In \figu\ref{fig|SersicIndexRedshift} we show the S\'{e}rsic indices $n$ (left) and effective radii \re\ (right) of galaxies with stellar mass $11.3<\mstare/\msune<11.7$. The latter are extracted from three (independent) galaxy samples, SDSS, COSMOS, and CANDELS, at broadly complementary redshift intervals $z<0.3$, $0.3<z<1$, and $z>0.5$, respectively, as measured by \citet{Bernardi17}, \citet{Huertas13a}, and \citet{Huertas15}. When taken as a whole, the data tend to suggest a relatively strong evolution of both the median $n$ and \re\ with redshift, consistent with $\propto (1+z)^{-1}$. Note that the observed redshift evolution in effective radius and S\'{e}rsic index at fixed stellar mass reported in \figu\ref{fig|SersicIndexRedshift} is expected to be contributed by different galaxies at different epochs, thus the actual evolution along the single galaxies may be even stronger (see discussion and references in \sect\ref{subsec|Progenitors}).
Also, due to the limited statistical range in the observational samples, the selections in \figu\ref{fig|SersicIndexRedshift} are performed at fixed stellar mass \emph{without} any cut in effective radius, as instead applied in \figus\ref{fig|GammaEvolutionDeVauc} and \ref{fig|GammaEvolutionSer}. However, we have carefully checked that considering redshift evolution only at fixed stellar mass in the \citet{Sonne13} data would, if anything, induce an even stronger flattening of \gamm\ with redshift, strengthening our conclusions on the necessity for a strong increase of the S\'{e}rsic index with cosmic time.
A varying S\'{e}rsic index with redshift for the most massive galaxies has not been fully confirmed by all observational groups. For example,
\citet{Ascaso13} find that $n$ only mildly increases from $z=0.9$ to $z=0.1$. However, it is possible to ascribe these apparently contrasting results to selections of different galaxy morphological types \citep[see, e.g.,][]{Huertas15}.

We conclude this \sect\ emphasizing that we have tested a number of physically and/or empirically motivated alterations
to the reference semi-empirical models presented above. For example, we have attempted to include an increasing gas fraction associated to the central galaxy back with cosmic time, as suggested by the data compilation of \citet{Stewart09}. We have assumed that the gas follows an exponential profile with half-mass radius twice the effective radius \citep[e.g.,][and references therein]{Guo11}. The resulting impact on \gamm\ is negligible, mainly because the gas fractions competing to massive galaxies is overall quite low \citep[e.g.,][]{Stewart09}.

\subsection{Probing variations in the stellar Initial Mass Function}
\label{subsec|EvolutionIMF}

In \sect\ref{subsec|EvolutionSersicIndex} we have highlighted that irrespective of the input stellar IMF, host halo mass, or even halo mass profile, a redshift-dependent S\'{e}rsic index is an essential component to induce a steep decrease in \gamm\ with increasing redshift, in line with observations.
In this \sect we specifically focus on the stellar IMF.

\begin{figure*}
    \center{\includegraphics[width=18truecm]{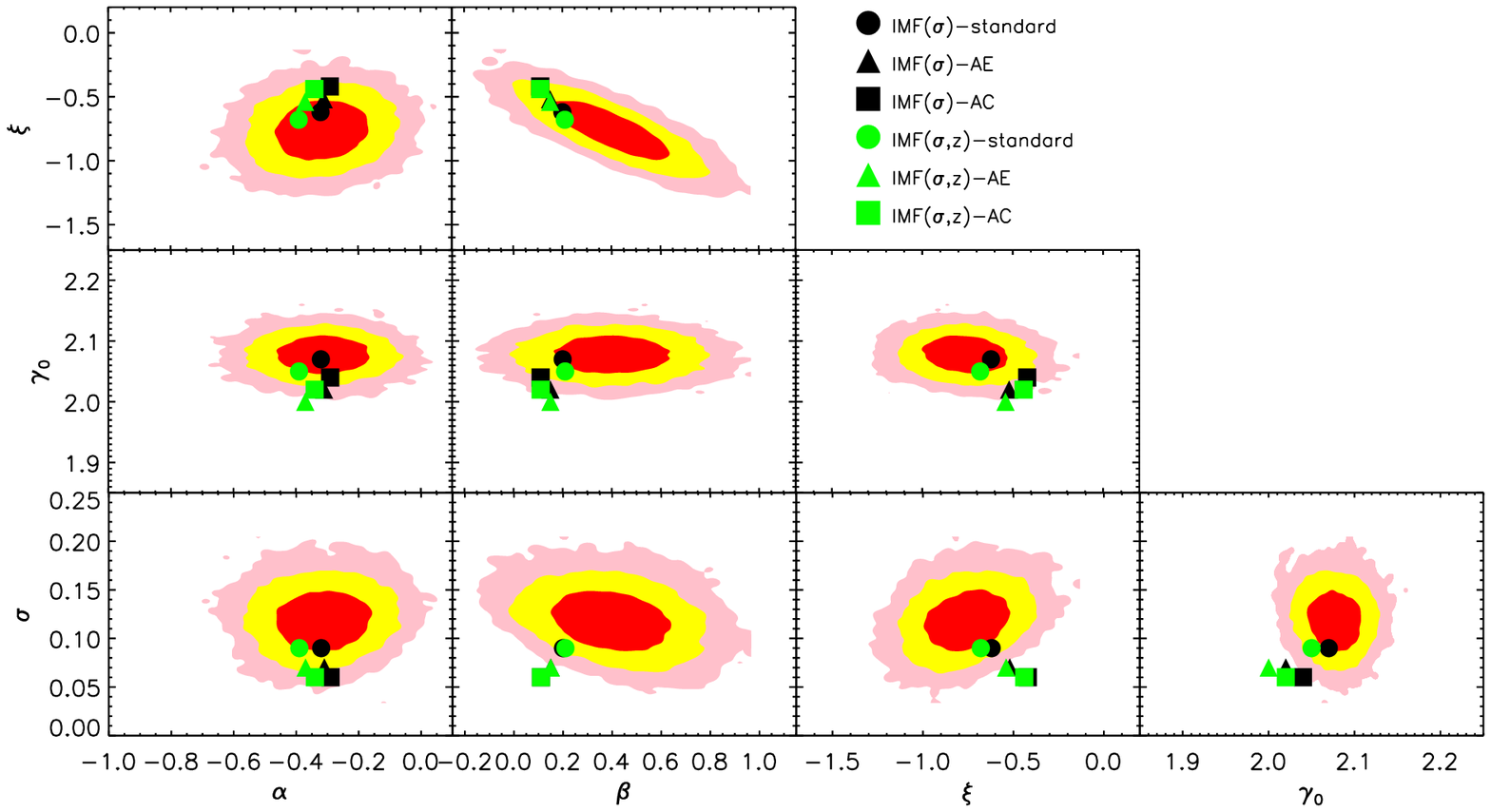}
    \caption{Similar format to \figu\ref{fig|GammaPosteriorProbIMF}. The red, yellow, and pink regions mark the 1$\sigma$, 2$\sigma$, and 3$\sigma$ contour levels for each pair of the parameter space extracted from the \citet{Sonne13} data (\eq\ref{eq|Sonne13gamma}). The data are compared with the predictions from the semi-empirical models based on a IMF variable with velocity dispersion (\eq\ref{eq|MLratioCappellari}) and a IMF variable in both velocity dispersion \emph{and} redshift (\eq\ref{eq|MLratioCappellariRedshift}; black and green symbols, respectively). The circles, triangles, and squares refer, respectively, to models with standard profiles, with adiabatic expansion, and with adiabatic contraction, as labelled. \label{fig|GammaPosteriorProbIMFz}}}
\end{figure*}

It is apparent from both \figus\ref{fig|GammaEvolutionDeVauc} and \ref{fig|GammaEvolutionSer}, that at $z\lesssim 0.1$ a Chabrier IMF (long-dashed, red lines) tends to be disfavoured with respect to more bottom-heavy IMFs, in line with \papI. This is more quantitatively outlined in \figu\ref{fig|GammaPosteriorProbIMF} which reports the covariances on the parameters of \eq\ref{eq|Sonne13gamma}, measured by \citet{Sonne13}. The red, yellow, and pink regions bracket, respectively, the 1$\sigma$, 2$\sigma$, and 3$\sigma$ contour levels for each pair of the four parameters $\xi$, $\alpha$, $\beta$, $\gamma_0$, and $\sigma_{\gamma'}$. Following \papI, we apply the same fitting procedure used in \citet{Sonne13}, modelling each of our mock galaxy samples as a Gaussian distribution with a mean given by \eq\ref{eq|Sonne13gamma} and with dispersion $\sigma_{\gamma'}$. The fit produces a posterior probability distribution function for the parameters $\gamma_0$, $\alpha$, $\beta$, $\xi$ and $\sigma_{\gamma'}$. For each model realization we plot the peak of the posterior probability distribution function in \figu\ref{fig|GammaPosteriorProbIMF}. The circles, triangles, and squares refer, respectively, to models with standard profiles, with adiabatic expansion, and with adiabatic contraction, as labelled.

\figu\ref{fig|GammaPosteriorProbIMF} confirms the results of \papI: even on the assumption of a S\'{e}rsic \emph{intrinsic} stellar profile and \devac\ as ``observed'' profile, a Chabrier IMF continues to be highly disfavoured with respect to a more bottom-heavy IMF, such as the variable one given in \eq\ref{eq|MLratioCappellari}.
Uncontracted NFW dark matter profiles also continue being highly preferred by the data. Adiabatic contraction tends in fact most notably to decrease $\beta$, weakening the dependence on stellar mass, and also to appreciably decrease the intrinsic scatter $\sigma$. Adiabatic expansion (black triangles) tends instead to progressively lower the dark matter contribution in the inner regions allowing for a more dominant role of the steeper stellar component, thus increasing the zero point $\gamma_0$, increasing $\beta$, and decreasing $\xi$.

Even more interestingly, as hinted at in \sect\ref{subsec|EvolutionSersicIndex}, a Chabrier IMF seems gradually more in line with the data at $z\gtrsim 0.7$ (long-dashed, red line in \figu\ref{fig|GammaPosteriorProbIMFz}). We quantify this trend in \figu\ref{fig|GammaPosteriorProbIMFz}. In the specific, we here compare two models, a variable IMF as given in \eq\ref{eq|MLratioCappellari} constant in time (black symbols), and a model characterized by an IMF variable in both velocity dispersion and time as in \eq\ref{eq|MLratioCappellariRedshift}. \figu\ref{fig|GammaPosteriorProbIMFz} clearly shows that, while the data may continue disfavouring adiabatic contraction and expansion, at present they cannot distinguish between a time-constant or time-variable IMF, being both consistent at the $\sim 1\sigma$ level in the full space of parameters.

There are currently few measurements of the evolution in the stellar IMF that we can compare our model to. \citet{Shetty14} used stellar dynamics
to put an upper limit on the IMF of a sample of 68 massive galaxies at $z\sim0.8$, finding that, on average, the IMF cannot be heavier than a Salpeter IMF.
\citet{Sonne15} used strong lensing and stellar kinematics to constrain the evolution in the stellar IMF normalization of the lenses in the \citet{Sonne13} sample. They found a slight preference for an IMF that becomes heavier with time \citep[parameter $\zeta_{\mathrm{IMF}}$ in ][]{Sonne15}.
\citet{Martin15a} constrained the IMF slope of a set of 48 massive galaxies at $0.9 < z < 1.5$ using spectroscopic measurements of absorption lines sensitive to the abundance of low-mass stars. They found similar IMF slopes compared to samples of galaxies at $z\sim0$.
All in all, current data may still not be sufficient enough to constrain the degree of evolution in the stellar IMF along cosmic time.

A possible way to break this degeneracy could rely on probing the evolution in the normalization of the Faber-Jackson relation \citep[e.g.,][]{FJ} and/or of the fundamental plane of early-type galaxies \citep[e.g.,][]{Dressler87}. As discussed by a number of groups \citep[e.g.,][]{Prugniel97,Borriello03,Mamon05,Mamon05b},
the central velocity dispersion $\sigma$ of ellipticals is heavily dominated by the stellar component. For a one-component (stars only) spherical isotropic galaxy model with S\'{e}rsic profile, the virial theorem can be written as $G\mstare/\ree=K(n)\sigma^2$, where $K(n)$ is a dimensionless coefficient depending only on the S\'{e}rsic index $n$. Taking into account that massive, bulge-dominated SDSS galaxies approximately follow $\mstare\propto \sigma^2$ for $\sigma\gtrsim 100$ km/s \citep[e.g.,][]{Barausse17,Shankar17BH}, one can then express the ratio $\mstare/\sigma^2$ as
\begin{equation}
\frac{\mstare}{\sigma^2}\propto K(n[z])\ree(z)\, .
    \label{eq|virialz}
\end{equation}
Assuming that the S\'{e}rsic index evolves with redshift as $n[z]\propto(1+z)^{-1}$, it would imply a decrease from $n\sim 6$ to $n\sim 3$ from $z\sim0$ to $z\sim 1$ for the galaxies of interest here, according to what reported in the left panel of \figu\ref{fig|SersicIndexRedshift}. Table 4 in \citet[e.g.,][]{Prugniel97} shows that $K(n)$ would increase from $K(n)\sim 3$ when $n\sim 3$, to $K(n)\sim 5.8$ when $n\sim 6$, implying a growth with redshift $K(n[z])\propto (1+z)^{0.96}$ which would balance the decrease in redshift of the effective radius, $\ree\propto (1+z)^{-1}$ (right panel of \figu\ref{fig|SersicIndexRedshift}) making the ratio $\mstare/\sigma^2$ about constant in time \emph{if} there is no additional time-evolution in $\mstare/L$.

\begin{figure}
    \center{\includegraphics[width=8.5truecm]{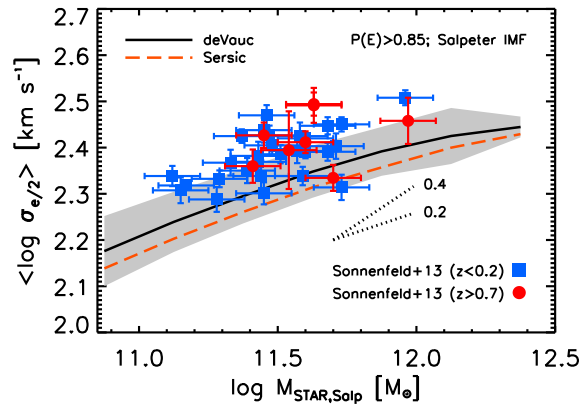}
    \caption{Line-of-sight velocity dispersion averaged within a circular aperture of radius equal to half of the effective radius $\sigma_{\rm e/2}$
    as a function of (Salpeter) stellar mass for for our sample of SDSS galaxies, assuming a \devac\ (red long-dashed line) and S\'{e}rsic (black solid line) light profiles.
    The grey shaded region marks the 1$\sigma$ dispersion around the \devac\ curve.
    The blue squares and red circles are, respectively, the $z<0.2$ and $z>0.7$ \citet{Sonne13} data points. There is no apparent segregation between the latter two
    subsamples. \label{fig|SigmaMstar}}}
\end{figure}

\figu\ref{fig|SigmaMstar} shows the velocity dispersion-stellar mass relation for our SDSS sample of galaxies. Velocity dispersions
are rescaled to a circular aperture of radius equal to half of the effective radius $\sigma_{\rm e/2}$. The long dashed orange and solid black lines refer to the
\devac\ and S\'{e}rsic stellar masses, respectively. The SDSS results are compared to the $z<0.2$ and $z>0.7$ \citet{Sonne13} data, shown as
blue squares and red circles, respectively. The \citet{Sonne13} data do not show any evident redshift evolution in the $\sigma(\ree/2)-\mstare$ plane.
The slight offset between the SDSS and the \citet{Sonne13} data has already been noted in \papI. As discussed there, a substantial part of the offset is most probably arising from a combination of a $\sim 0.1$ dex higher mass-to-light ratios and larger apertures (larger effective radii) adopted by \citet{Bernardi13} and \citet{Bernardi17} with respect to previous works \citep{Hyde09a}. Indeed,
\citet[][see also \citealt{Treu06}]{Auger10} did not find any noticeable offset between SLACS and \citet{Hyde09a}.
Nevertheless, one could still conceive the presence of some selection effects in strong lensing-selected early-type galaxies, being for example more
compact at fixed stellar mass. We verified, however, that by selecting in our SDSS-based mocks only those galaxies above the median $\sigma$-$\mstare$ in \figu\ref{fig|SigmaMstar},
our predictions are unchanged, as expected given that at any redshift we are selecting galaxies at fixed stellar mass and effective radius.

Recent work by \citet[][see also, e.g., \citealt{Shetty14}]{Shetty15} has also shown the $\sigma(\ree)-M(<\ree)$ relation to be constant up to $z\sim 1$ for galaxies with (dynamical) mass above $M(<\ree)>2\times 10^{11}\, \msune$.
\citet[][and references therein]{Beifiori17} analysing 19 intermediate-mass red-sequence galaxies at $z\sim 1.5$ from the KMOS Cluster Survey, on the other hand found evidence for some evolution in the $\sigma(\ree)-M(<\ree)$ relation since $z\sim 0$. Clearly the relative roles of the $\mstare/L$ ratio against, e.g., dark matter fraction, structure, non-homology, or anisotropy in shaping the massive end of the $\sigma-\mstare$ relation at different cosmic epochs are still debated. More detailed and focused modelling, which is beyond the scope of the present paper, will further discuss this and other related issues in separate work (Shankar et al. in prep.). Irrespective of all of the above, an increasingly lighter IMF at higher redshifts would be anyway insufficient, by itself, to explain the flattening of \gamm\ at earlier epochs (see \figu\ref{fig|GammaEvolutionSer}).

\subsection{Dark Matter fractions as a function of redshift}
\label{subsec|DMfracRedshift}

\begin{figure}
    \includegraphics[width=8.5truecm]{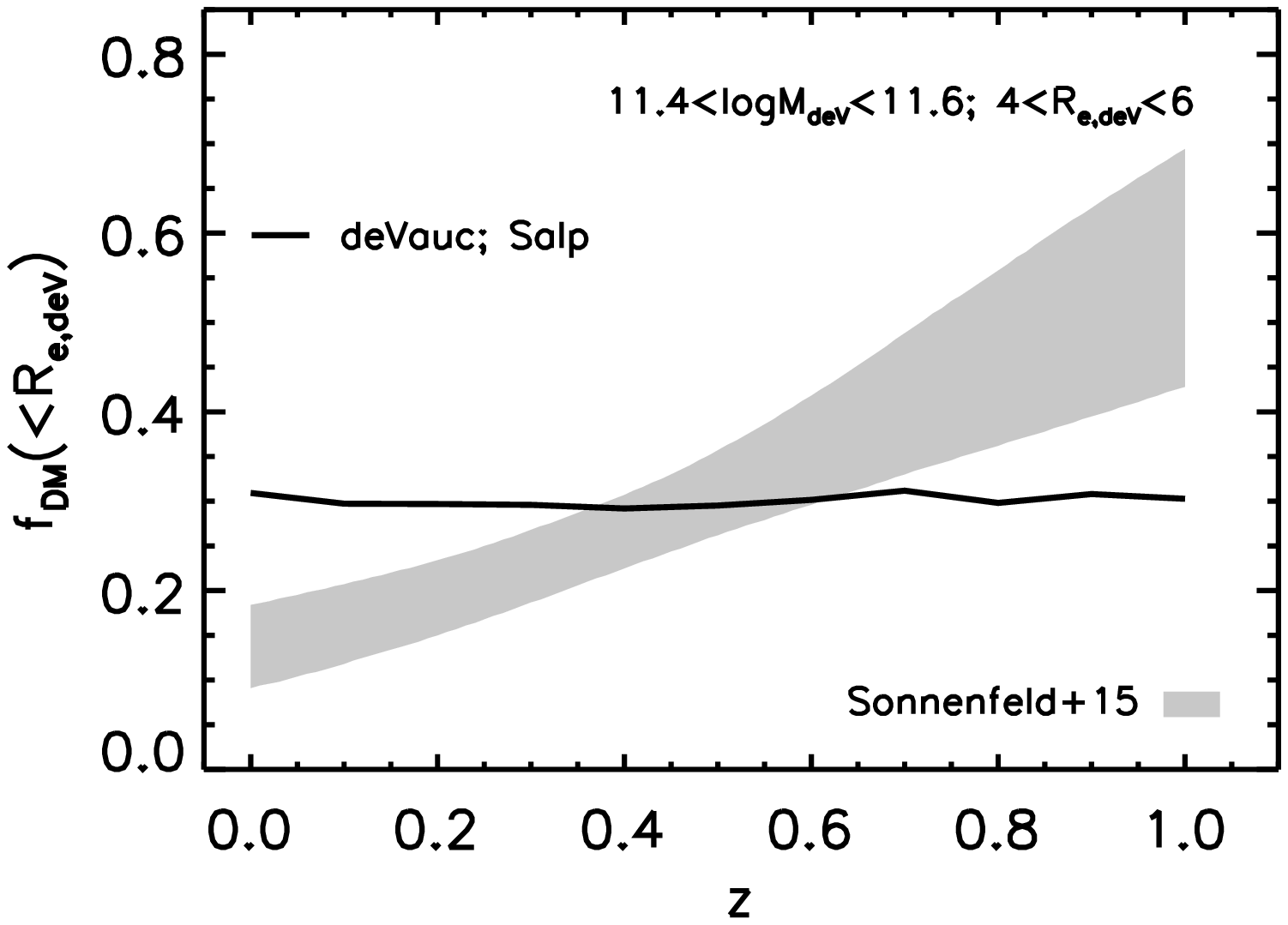}
     \caption{Predicted projected dark matter fractions within the effective radius as a function of redshift for galaxies with (\devac) stellar mass $11.4<\log \mstare/\msune <11.6$ and effective radius $4<\ree/\kpce<6$. The grey band are measurements for galaxies in the same interval of mass extracted from \citet{Sonne15} sample, assuming NFW and \devac\ profiles for the dark matter and the stellar components, respectively. The model fails in properly reproducing the redshift increase of the dark matter fraction. \label{fig|fDMz}}
\end{figure}

For completeness, in \figu\ref{fig|fDMz} we present the predictions for the redshift evolution of the \emph{projected} dark matter fraction $f_{\rm DM}(<R)=\mhaloe(<R)/[\mstare(<R)+\mhaloe(<R)]$ computed at the (\devac) effective radius $R_{\rm e, deVauc}$.
At all redshifts we then compare the select in our mock catalogues galaxies with (\devac) stellar mass $11.4<\log \mstare/\msune <11.6$ and effective radius $4<\ree/\kpce<6$, fixed with redshift.

Dark matter fractions for the \citet{Sonne13} sample of lenses have been measured by \citet{Sonne15}.
These measurements have been obtained by fitting two component (bulge+halo) models to the observed Einstein radius and stellar velocity dispersion, i.e.,
the same data used to measure the total density slope \gamm.
\citet{Sonne15} made the following assumptions: isotropic orbits, spherical symmetry,
a dark matter distribution following an NFW profile with scale radius fixed to ten times the effective radius, and a stellar distribution following a spherically-deprojected \devac\ profile with a constant mass-to-light ratio. For consistency, we compare these measurements with models with fixed $n=4$ at all redshifts.

As \figu\ref{fig|fDMz} shows, the model predicts a more or less constant dark matter fraction with time, while the data indicate an increase of dark matter fractions at higher redshift. This suggests that a model with a fixed \devac\ profile may not be a good description of reality, as the comparison with the evolution in \gamm\ already suggested.

It is however important to realize that the determination of \gamm\ is a much more robust measurement than $f_{\rm DM}$.
The former has in fact a very mild dependence on the assumed stellar
profile being determined by fitting a power-law density profile to the observed velocity dispersion
and Einstein radius. Varying the stellar density profile has thus only a mild effect on the inference, since it only affects the
distribution of the kinematical tracers, which has a small effect on the model velocity dispersion. On the other hand,
modifying the stellar profile has a much more significant
impact on the measurements of $f_{\rm DM}(<R)$, because it also changes the
distribution of mass of the model, greatly affecting both the velocity dispersion and the Einstein radius.

The specific choice of host halo mass in the mocks clearly plays a
substantial role in the predicted $f_{\rm DM}(<R)$. For example, we have verified that
assuming at $z\lesssim 0.1$ a host halo mass a factor of $\lesssim 3$ lower than the one
currently adopted in our reference mocks, and gradually increasing at higher redshifts, tends to improve the match to the data.
A lower value of the host halo mass in the
local Universe may be accounted for by a substantially steeper $\mstare-\mhaloe$ relation,
as recently suggested by the dynamical modelling in SDSS galaxies by \citep[][see also \citealt{Shankar06}]{Bernardi17b}.
Irrespective of this, we have verified that including in the \devac, Salpeter model
(solid line in \figu\ref{fig|GammaEvolutionDeVauc}) a varying host halo mass as the one
required to align the model with the \citet{Sonne15} $f_{\rm DM}(<R)$,
would imply a \gamm$\sim 2.20$ at $z\lesssim 0.1$ gradually increasing to \gamm$\sim 2.24$ at $z\sim 1$,
still much steeper than what suggested by the strong lensing data for which \gamm$\sim 2.05\pm0.08$ at $z\sim 1$
(grey area in \figu\ref{fig|GammaEvolutionDeVauc}).

\section{Discussion}
\label{sec|discu}

\subsection{Impact of observational selection effects}
\label{subsec|biases}

The main point emphasized in \sect\ref{sec|Results} is that semi-empirical models struggle to reproduce the
apparent increase of \gamm\ as a function of cosmic time, \emph{unless} the S\'{e}rsic index $n$ is allowed to strongly vary as $\propto (1+z)^\delta$, with $\delta \lesssim -1$. To set these conclusions on a firmer foot, it is important to properly understand whether some selection effects may be biasing the apparent
redshift evolution of \gamm\ and in turn our physical interpretation of the observational data.

\begin{figure*}
    \center{\includegraphics[width=18truecm]{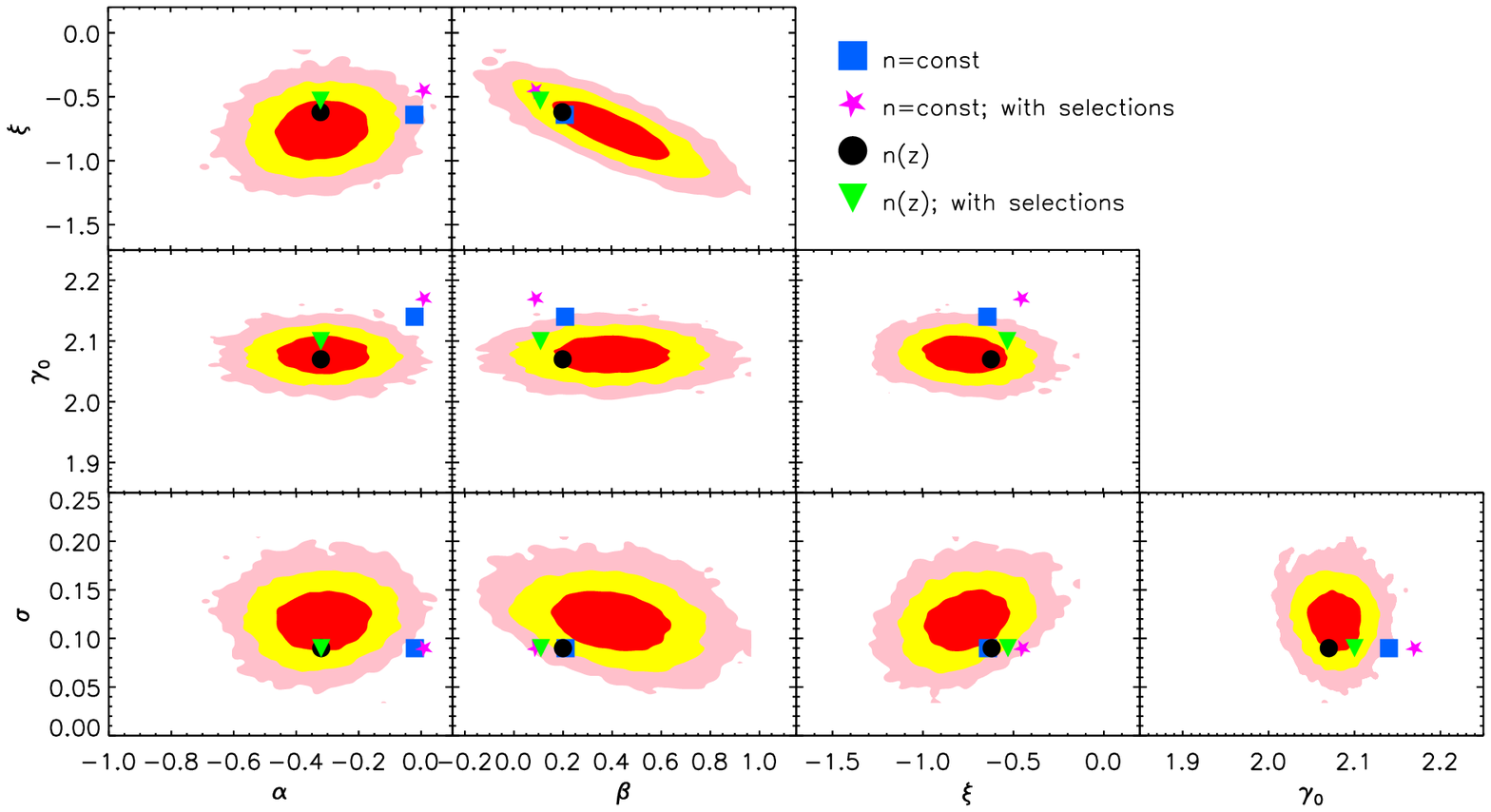}
    \caption{Similar format to \figu\ref{fig|GammaPosteriorProbIMF}. The red, yellow, and pink regions mark the 1$\sigma$, 2$\sigma$, and 3$\sigma$ contour levels for each pair of the parameter space extracted from the \citet{Sonne13} data (\eq\ref{eq|Sonne13gamma}). The data are compared with predictions from the semi-empirical models based on a IMF variable with velocity dispersion (\eq\ref{eq|MLratioCappellari}), and, as labelled, a constant S\'{e}rsic index with (magenta stars) and without (blue squares) strong lensing observational selection effects, and a redshift-dependent S\'{e}rsic index with (green triangles) and without (black circles) strong lensing observational selection effects. See text for details. \label{fig|GammaPosteriorProbSelections}}}
\end{figure*}

The main effect of strong lensing selection on a sample of galaxies is to skew its distribution towards a region of parameter space corresponding to galaxies with a higher strong lensing cross-section. This can in principle alter the distribution in \gamm\ of a sample of strong lenses compared to that of the general population of galaxies. To assess the importance of this effect, we carry out the following experiment. For each galaxy in our mock, we calculate its strong lensing cross-section, i.e., the area enclosed within its radial caustic in the source plane, assuming spherical symmetry. This requires assuming a source redshift, $z_s$. We fix $z_s=0.6$ for galaxies at redshift $z < 0.3$ and $z_s=2$ for galaxies at a higher redshift, to mimic the source redshift distribution of SLACS and SL2S lenses, on which the \citet{Sonne13} measurements are based.
We then draw a subset of 10,000 objects from the mock, weighting each object by its strong lensing cross-section. This subsample is intended to resemble a set of strong lensing-selected objects. Finally, we fit our model distribution of \gamm\ to this mock strong lens sample and compare it with the result obtained over the whole sample.

Our results are reported in \figu\label{fig|GammaPosteriorProbSelections}. The latter compares the \citet{Sonne13} 1$\sigma$, 2$\sigma$, and 3$\sigma$ contour levels of the posterior probability of each data pair in \eq\ref{eq|gamma}, with the predictions from our reference semi-empirical models based on a IMF variable with velocity dispersion (\eq\ref{eq|MLratioCappellari}), and a variable $n\propto(1+z)^{-1}$ S\'{e}rsic index, inclusive (green triangles) or not (black circles) of strong lensing selection effects.
We find small differences. The more relevant effect is a $\sim 0.03$ increase in the average slope, $\gamma'_0$ but with virtually no effect on its dependence on redshift, i.e., on $\alpha$. This is well understood, as galaxies with a steeper density profile have a larger lensing cross-section. However, the shift with respect to the distribution of the whole sample is small, owing to the relatively small scatter in \gamm. For completeness, in the same \figu\ we also include the results of the model with a constant S\'{e}rsic index. The latter naturally predicts (blue squares) a null evolution in redshift, thus $\alpha\approx 0$, and thus a much larger value of $\gamma'_0$. Including the above-quoted strong lensing selection effects (magenta stars) if anything further worsens the match to the data, further strengthening our conclusions on the need for a strong evolving S\'{e}rsic index.

\citet{Xu17}, have also stressed that additional systematics in the comparison between models and data could arise from varying degrees of stellar anisotropy, which should anyhow be limited to differences of $\lesssim 0.1$ in \gamm\ between $z=0$ and $z=1$, comparable to, or even smaller than, current observational uncertainties.

\subsection{Comparison with Semi Analytic Models}
\label{subsec|SAMs}

Up to this point we have compared available data with basic semi-empirical models to minimize the number of input parameters and assumptions, thus allowing for a more transparent and secure interpretation of the data. We have also calculated the redshift evolution in the total mass density profiles
of massive galaxies extracted from the \citet{Guo11} semi-analytic model. To this purpose
we used their publicly available online catalogues\footnote{http://www.g-vo.org/MyMillennium3} with stellar and halo masses, and applied the same virial radii, halo profiles, and concentrations as in our reference models. We also compared with the \citet{Shankar13} variant of the \citet{Guo11}, which includes gas dissipation in major mergers and parabolic orbital energies.

Analogously to our mocks, we find these semi-analytic models to predict for bulge-dominated, massive galaxies a steeper evolution of \gamm\ out to $z\sim 1$ with respect to the \citet{Sonne13} data.
In line with our previous findings, modelling galaxy profiles as S\'{e}rsic with redshift-dependent S\'{e}rsic index $n$ provides an improved match to the data.

\subsection{Evolution along the ``progenitors''}
\label{subsec|Progenitors}

So far we have always discussed structural evolution at \emph{fixed} stellar mass and size.
This may not help us to distinguish among different evolutionary scenarios.
In fact, the impact of new galaxies entering the selection at later epochs may substantially contribute
to the observed structural evolution of massive galaxies \citep[e.g.,][and references therein]{Cimatti12,Carollo13,Shankar15},
the so-called ``progenitor bias'' effect \citep[e.g.,][]{van96,Saglia10}.
So, what would be the evolution of the total mass density profiles along the ``progenitors''?
Of course, the galaxy stellar mass is expected to be lower in the progenitors, possibly inducing flatter \gamm\ values.

It is clearly observationally challenging to identify galaxy progenitors along cosmic time. A number of techniques have been proposed in the last decade to achieve this goal, most notably number density conservation \citep[e.g.,][]{vandokkum10,Patel13,Huertas15}.
\citet{vandokkum10}, in particular, suggest that massive galaxies, in the range of stellar mass of interest here, should decrease in mass at earlier redshifts as
\begin{equation}
\log \mstare[z]=\log \mstare[z=0]-0.67\log(1+z) \, .
\label{eq|Mstarz}
\end{equation}
This would imply progenitors $\sim 0.2$ dex lower at $z\sim 1$, in broad agreement with other independent observationally-driven studies \citep[e.g.,][]{Lidman12,Vulcani16}, and semi-empirical modelling \citep[e.g.,][]{Shankar14,Shankar15}.

We start with SDSS galaxies in our sample with $\log \mstare\sim 11.5$ (Salpeter), characterized by median (\devac) effective radii of $\ree\sim 8$ kpc. We then evolve these galaxies back to $z=1$, assuming an evolution in stellar mass as given by \eq\ref{eq|Mstarz} and in effective radius of $\propto(1+z)^{-1}$.
For these ``progenitor'' galaxies the resulting total mass density slope is \gamm$=2.03\pm0.08$, adopting the posterior probability distribution by \citet{Sonne13} data and \eq\ref{eq|gamma}.
Our reference semi-empirical model with intrinsic S\'{e}rsic profile and observed \devac\ for the stellar component, with a redshift evolution in both the S\'{e}rsic effective radius and S\'{e}rsic index as $\propto(1+z)^{-1}$, would predict for the same set of progenitor galaxies \gamm$\sim 2.00$, in full agreement with what inferred from the strong lensing posterior probability distribution. On the other hand, assuming no evolution in the S\'{e}rsic index would yield \gamm$=2.31$, more than $\gtrsim 3\sigma$ away from the strong lensing data. A strong evolution in the S\'{e}rsic index thus seems to remain a necessary condition in evolving the single galaxies, and not a simple byproduct of progenitor bias effects.

But how can the S\'{e}rsic index evolve so strongly with cosmic time? An increase with S\'{e}rsic index has been suggested as a consequence of mergers.
\citet{Hop08FP} suggested that at each merger event the S\'{e}rsic index should increase by an amount
\begin{equation}
\Delta n \sim k \left(\frac{\Delta M}{M_{\rm cen}}\right)
\label{Eq|mergerSersic}
\end{equation}
where $k$ is a constant and $\Delta M$ is the amount of stellar mass increase in the central galaxy $M_{\rm cen}$ after a merger.
\citet{Hop08FP} suggested that $k\sim 1$, while \citet{Hilz13} found that for galaxies embedded in dark matter haloes, minor mergers induce a significantly faster size growth and profile reshaping with respect to galaxies without dark matter \citep[see also][]{Nipoti03}. Forcing \eq\ref{Eq|mergerSersic} to broadly match the \citet{Hilz13} numerical simulations (their \figu5), would roughly correspond to $k\sim 2-3$ for major mergers and $k\sim 5-20$ for minor mergers,
which would imply an increase of S\'{e}rsic index of up to five units for a stellar mass growth of just a factor $\sim 1.6$ (solid, red line in their \figu5).
It is highly unclear whether any of the latter proposed solutions will be able to yield an increase in S\'{e}rsic index as strong and sharp as the one actually observed (\figu\ref{fig|SersicIndexRedshift}). A close inspection of this important issue is beyond the scope of this paper and it will be explored in future work.

\subsection{Comparison to previous work on the topic}
\label{subsec|PreviousWork}

One of the most novel results of this work is the interpretation of the redshift evolution of \gamm.
The observed evolution of a decreasing \gamm\ at increasingly higher redshifts, cannot simply be due to only selection effects \citep[e.g.,][]{Sonne13,Sonne15}
or systematic measurement errors. Independent
observations point to a similar, if not stronger, flattening of \gamm\ with increasing redshift \citep[e.g.,][]{Dye14,Mori14}.
Previous observational work also inferred a less or equally significant flattening of the total mass density profiles back in cosmic time
\citep{Koop06,Ruff11,Bolton12}.

\citet{Ruff11} suggested that dissipative processes might have played some role in the growth of massive galaxies and the steepening of the mass profile since $z \sim 1$.
\citet{Bolton12} noted that significant baryonic dissipation at late times is in contrast with the very low star formation rate and old stellar populations characterizing massive, early-type galaxies. Inspired by the numerical simulations of \citet{Nipoti09}, they instead pointed out the possibility that off-axis major dry mergers could contribute towards the observed evolution in \gamm.

However, purely dry mergers have been disfavoured by \citet{Sonne14merge} as a viable mechanism for the observed trend of \gamm\ with redshift, as
they pointed out that such mergers produce a flattening of the density slope with time, once cosmologically motivated stellar-to-halo mass ratios and self-consistent definitions of \gamm\ between simulations and observations are used.
\citet{Sonne14merge} suggested that mergers with small amounts of dissipation and nuclear star formation (consistent with observational constraints) are required to fit the lensing data.

Cosmological hydro-simulations dedicated to the full study of the evolution of massive, early-type galaxies, have also been used to made predictions on the evolution of the density slope.
\citet{Dubois13} find an average \gamm\ that becomes steeper with time, in agreement with the data.
\citet{Joha12}, on the other hand, find a nearly flat evolution for the total logarithmic density slope below $z \lesssim 1$, and steeply rising above $z\sim 1$.

\citet{Xu17} have recently carefully analyzed the total mass density slopes of early-type galaxies in the Illustris simulation finding a mild increase towards lower redshifts, in broad agreement with the data.

\citet[][]{Remus17} further extended their previous analysis \citep{Remus13} presenting evidence from their cosmological hydrodynamical simulations that the dark matter fraction within the half-mass radius is lower at higher redshifts, in line with what discussed here and the obervationally-oriented results by \citet{Tortora14b}. On the other hand, they find their total mass density slopes $\gamma$ to behave in stark contrast to observations, predicting steeper slopes at higher redshifts. They claim this discrepancy between observations and simulations to be mostly a result of the observational methodology adopted to determine the density slopes. In our semi-empirical models we do not find evidence for such strong tensions with the data. Indeed, at fixed stellar mass and effective radius our \gamm\ slopes are already predicted to be roughly constant with redshift. Once accounting for a S\'{e}rsic index evolution as suggested by independent observations, the \gamm\ are then predicted to naturally steepen with cosmic time, in nice agreement with \citet{Sonne13}.

\section{Conclusions}
\label{sec|Conclu}

In this second paper of the series we carry out a systematic investigation of the total mass density profile of massive ($\log \mstare/\msune \sim 11.5$) early-type galaxies and its dependence on redshift, specifically in the range $0\lesssim z \lesssim 1$. We start from a large sample of SDSS early-type galaxies with measured stellar masses and effective radii assuming two different intrinsic profiles, \devac\ and  S\'{e}rsic. Via abundance matching relations we assign to galaxies dark matter haloes with standard $\Lambda$CDM profiles and concentrations. We then compute the total, mass-weighted density slope at the effective radius \gamm, and study its redshift dependence at fixed stellar mass and effective radius. We find that a necessary condition to induce an increasingly flatter \gamm\ at higher redshifts, as suggested by current strong lensing data, is to allow the intrinsic stellar profile of massive galaxies to be S\'{e}rsic and the input S\'{e}rsic index $n$ to vary with redshift as $n(z)\propto (1+z)^{\delta}$, with $\delta \lesssim -1$. This conclusion holds irrespective of the input \mstar-\mhalo\ relation, the assumed stellar initial mass function, or even the chosen level of adiabatic contraction in the model. Secondary contributors to the observed redshift evolution of \gamm\ may come from possible increases at earlier times in the adiabatic contraction and/or contributions from a bottom-light IMF. We have also explored the possible impact of strong lensing selection effects. The latter may induce an increase in the average slope, $\gamma'_0$, but have virtually no effect on its dependence on redshift, i.e., on the parameter $\alpha$ in \eq\ref{eq|Sonne13gamma}.
A steadily increasing S\'{e}rsic index with cosmic time is supported by independent observations, though it is not yet clear whether cosmological hierarchical models (e.g., mergers) are capable of reproducing such a fast and sharp evolution.

\section*{Acknowledgments}
FS acknowledges valuable discussions with A. Beifiori.
We thank the referee for a number of very useful comments and suggestions that substantially improved the presentation of our results.

\bibliographystyle{mn2e_Daly}
\bibliography{../../../refMajor_Rossella}

\label{lastpage}
\end{document}